# Ordinary Low Alpha Proportional Counter with Low Cost Commercial Data Acquisition System

Tung Yuan Hsiao[1,*], C. H. Chen[1], Huan Niu[1]

[1]*Accelerator Laboratory, Nuclear Science and Technology Development Center, National Tsing Hua University, Hsinchu 30013, Taiwan*

**Abstract**

In this study, we present a low cost and commercially available data acquisition system (DAQ) for an ordinary low alpha proportional counter. By employing this DAQ system and aid of a simple physical model, we can easily rule out the common disadvantage of proportional type low alpha counters, such as sensitive to electromagnetic interference and vibration. The obtained results demonstrated that this method has improved the capability of an ordinary low alpha counter and even makes it easier to operate in a worse ground-loop laboratory.

## 1. Introduction

Over the last two decades, the sensitivity of electronics to ionization radiation has been of increased interest as ionization processes dominate the soft-error rate (SER) of electronics, for example, a type of error where a state of storage elements was changed. There are two primary radiation sources that cause SER. One is alpha-particles from radioactive contaminants in the chips, and the other one is cosmic rays. The errors caused by radioactive contaminants is mainly due to the purity of materials, which used in chip fabrication. Therefore, low background and highly sensitive alpha particle detection system are required to measure the radiation level of the material. In general, a commercially available 1000 $cm^2$ proportional counter can achieve a detection limit to 0.002 counts/$cm^2$/hour within three counts per hour background in 24 hours. Keep such a low background in long period needs to eliminate or suppress noise carefully. It's usually difficult for an old laboratory because of the complicated power ground loop, power usage, and ground vibration. Moreover, with the usage history, some alpha radioactive contaminants in the chamber of proportional counter getting increase due to the daughter nucleus of Radon and measured samples.

Generally, noise can be discriminated by a simple threshold level setting. However, the few high-energy spikes like noise could pass this discrimination to cause in noise level enhancement. This is a practical problem in low alpha counting. In traditional radiation measurement, a more complex data acquisition instrument, such as pulse shape analysis, is needed to separate signals from noise to solve this problem. Gordon et al., [1, 2] had published the excellent results of energy information of ultra-low alpha from XIA (ionization counter) [2, 3] by using the rise time and pulse height of active signal discrimination to check the ultra-low alpha event is contributed from sample or background for ionization counter.

From the Gordon report in CIRMS [4], the main disadvantage of the proportional counter for low alpha counting comes from electromagnetic interference (EMI) noise and vibration. Recently, we proposed a digital signal processing method as a data acquisition system for a large area proportional counter [5]. In this report, a homemade 241Am alpha source [6] has been detected by this system. At first, the pulse shape of the alpha particle has obtained and then the noise could be identified by comparing the recorded data with the alpha particle pulse shape. The results indicate that noise level could be decreased significantly for a large area proportional counter even if during high rate noise events.

## 2. Methods and Results

In this study, we use an ordinary alpha counter (Model 1950-1000), which bought from Alpha Science Inc (Now it is part of XIA LLC) [7]. As mention previously, this kind of alpha counter is sensitive to electromagnetic interference (EMI) noise and vibration. However, with the improvement of electronic devices and microprocessors, now we could have the chance to eliminate these influences on alpha counting. For Model 1950 alpha counter, the alpha detector window is 250 µg/cm$^2$ metalized Mylar, and outside dimensions are 46 cm wide, 55 cm deep and length of 11.4 cm. When a several MeV alpha particle emitted from a reaction, it's velocity is about $1 \times 10^7$ m/s.

The height of the Model 1950 alpha counter chamber is about 2 cm, and the height of the sample chamber is also 2 cm. When Alpha particles go through it, and to deposit in the counter chamber will take about few nanoseconds (3.3 ns). To compare this value with a typical rising time of preamplifier (Ortec 109PC) of a proportional counter is bellowed 1 microsecond (20 ns for Ortec 109PC). Therefore, the time for alpha particles traveling in the counter can be neglected. It means that all the electrons are created in a very short time (< 4 ns), and this time will not influence the rising time of the preamplifier. So, when the electrons travel to the anode wire and get signal multiplication, according to the electron velocity ($2 \times 10^6$ cm/s) and the height of counter (2 cm), it will take about one microsecond to reach anode wire. So, for the worst case and based on the above assumption, all charge in a counter should be collected in 1-2 μs, i.e., the rising time of preamplifier. It means in theoretically there is no more charge will be created after four ns, and after 1-2 μs, there is no more charge in the counter. According to the hypothesis, any non-exponentially decay signals should come from other circuit noise, something like EMI or vibration.

Figure 1 shows the schematic diagram of the system. According to the Shockley-Ramo theory on the induced charge [8], the charges are collected at an electrode to generate a current flow in the detector only. Thus a resistive feedback charge sensitive preamplifier integrating the induced charges as shown in Fig.1 [9]. Hence, the amplitude of the signal $V_{out}$ is given by

$$V_{out} = \frac{Q_f}{C_f} = \frac{\eta_{in} Q_t}{C_f} \qquad (1)$$

Where $\eta_{in}$ is the charge transfer efficiency, $Q_f$ is the charge accumulated on the feedback capacitor $C_f$, $Q_t$ is the total input charge, and $R_f$ is the feedback resistance of preamplifier. The maximum value of the voltage occurs at t = 0 and then pulse decays exponentially with a time constant equal to $R_f C_f$ [9]. Therefore, a true signal may be an exponentially decaying pulse signal, as shown in Fig. 1(a) afterward of the preamplifier. The other signals have not appeared as the expected and would be removed by our filter. For example, these phenomena of noise have shown in the Figs. 1(b) and Fig. 1(c). The Figs. 1(b) presents many random peaks and the Figs. 1(c) displays the shape of a spike only. Thus, we suspect such noises might be from the false signals of charge sensitive preamplifier and does not from the detector.

As mention above, the rising time is not significant in the proportional type alpha detector. Moreover, the decay time is as long as 400 μs. Therefore, the sampling rate we requisite is around 10 μs. In this study, we choose a Cortex-M3 RISC processor which equipped a nominal 1MSPS 12-bit ADC. For convenient, we used the Arduino DUE and a home mode ADC driver to match the signal condition. Theoretically, if we use 2byte to record the acquired data, the DAQ data rate is 2Mbyte per second, so we choose USB2.0 bus to transfer ADC data to the computer, it is fast enough to handle the data in situ. We have used the Raspberry Pi Model 3 B (computer), which equipped with 1.2GHz 64bit quad-core ARM Cortex-A53 and 1 G-Bytes LPDDR2 RAM. The Ortec 109PC preamplifier has been used along with Analog Devices AD8032A for matching the input impedance and ADC input voltage range. The schematic diagram in figure 1 shows that acquired events. As we mentioned above, we can quickly identify alpha and noise events.

We have mentioned that, we can identify the events and acquired from the DAQ system. Therefore, we employed this system in an uncontrolled environment (EMI and with vibration) to obtain the data. Then we acquire the background data for an hour and plotted, which has shown in Fig. 2. There are 21 events in hour data, and we can use an algorithm to identify the event, such as which one is alpha event and which one is noise event. From the results, it has been observed that, there are 4 alpha events (blue) and 17 noise events. These result demonstrations that, the DAQ system and algorithm are useful to rule out the unwanted events. Therefore, we attempted to use the DAQ system for 145 hours in the same environment.

Fig. 3 depicts the events counts vs. time plot, where the blue line is raw data, and the red line is filtered data. These results indicate that the noise events can be removed successfully from the acquired data. Moreover, these results exhibit a high noise events rate occurred during 130-145 hours, which may be caused by a compressor pump in the laboratory. Fig. 4 shows the counts vs. energy plot, which reveals that most of the noise arising from the low energy random noise events. Fig. 5 show the acquired data for homemade 241Am source, here we have employed the NIM module (Figure 5(a)) and DAQ system together to acquire data (Figure 5(b)). The utilized NIM modules are Canberra AFT Research Amplifier Model 2025, Canberra ADC 8701 and Series 35 plus MCA.

### 3. Discussions

Preamplifiers have been used when radiation is detected as a series of pulses of current, and these pulses are current flow into (or out of) the preamplifier. The feedback resistor $R_f$ (in parallel with the feedback capacitor) slowly discharges (resets) the feedback capacitor $C_f$, producing an exponential decay of each pulse with a time constant 50 μs. For this reason, the current pulses from the detector should be limited in duration of a few microseconds, as the longer pulses would have distortion due to the exponential decay. Figure 1(a) shows the ideal pulse shape acquired by the system. Figure 1(b) and 1(c) exhibit the random noise and spike noise, and these noises not shown the exponential decay, it infers that the signal doesn't come from the preamplifier. If the signal comes from the preamplifier, it must go through feedback resistor $R_f$ and feedback capacitor $C_f$, and it will obey physical nature and have an exponential decay behavior. Since these signals don't have this feature, it might be arising from the circuit after preamplifier. Therefore, we have employed this reason to rule out these noise signals from the acquired data. Fig. 2 shows the original raw data in one-hour, red ones are identified as random noise and spike noise, and blue ones are the signals, which behavior like exponential decay. Thus, it's clear to illustrate that, we could use this algorithm to rule out the noise and make the detector to work under noisy environment.

Figure 3 shows the measurement result for 150 hours, here the histogram in red color presents the original raw data, which doesn't remove the noise, and histogram in blue color presents the data, which remove the noise by using our algorithm. We can observe the lots of noise events around the collecting time of 130 to 150 hours. In the past, if appeared such noise, then we required to remove this period of data, and extend the measurement time or repeat the measurement. Now with this algorithm, we don't have to waste time to redo the measurement. Fig. 4 show the same data but plotted in a spectrum. Red color represents the original data, which doesn't remove noise, and blue one presents filtered data. We can see most noises came from the low energy region, and few came from high energy portion. The noise in the low energy region might be from EMI noise or vibration, and the noise in high energy area might be from the high voltage breakdown during the measurement.

Figure 5 shows the spectra of 241Am, which one was acquired by conventional NIM module and used the Series 35 plus MCA as shown in the histogram (Figure5(a)), the other was received by our system (Figure5(b)). It can be found that our method could remove the noise in either the low energy or high energy region and results in the spectrum more clearly observable. The data were measured at the same time by a 1-2 splitter.

## 4. Conclusions

In this study, we have demonstrated that with the aid of digital pulse shape discrimination, the noise could efficiently suppress, which arises from EMI or vibration. In the meantime, this method also could be applied to remove low energy noise from the spectrum and the modification of DAQ cost as low as 100 USD.

**Acknowledgment:** This work is supported by the Ministry of Science and Technology, Taiwan under grants no. MOST-106-2119-M-266-001 and MOST-105-2119-M-266 -001. We also thank Dr. Srinivasu Kunuku for his valuable inputs.

**FIGURE CAPTIONS:**

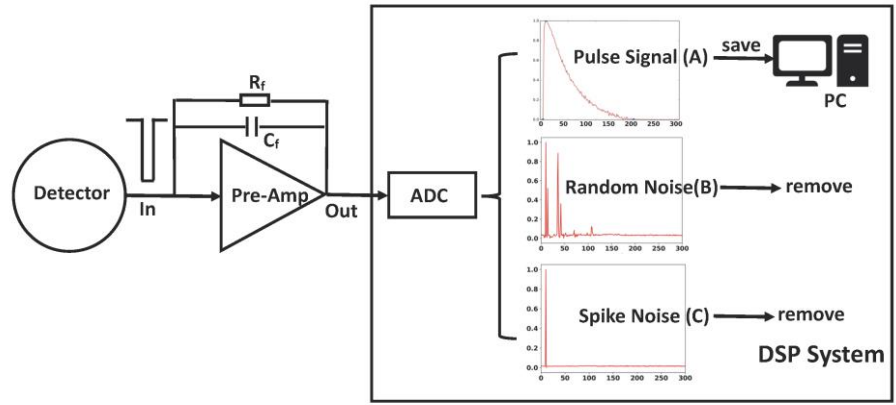

Figure 1. The resistive feedback charges sensitive preamplifier.

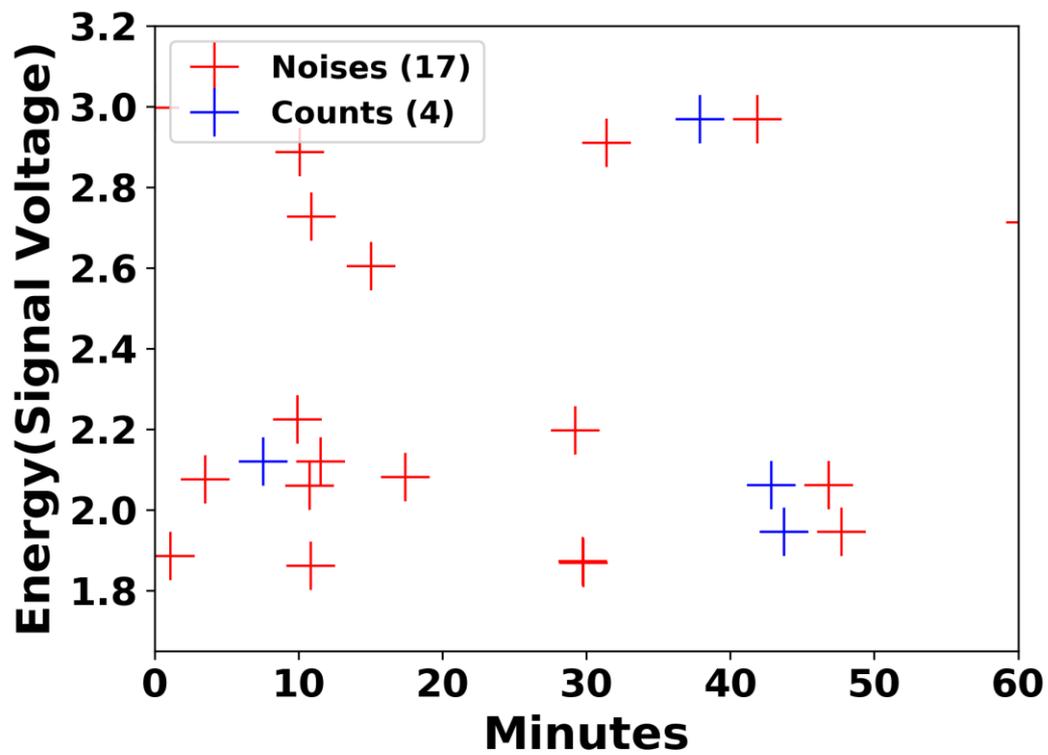

Figure 2. The counting result in one hour.

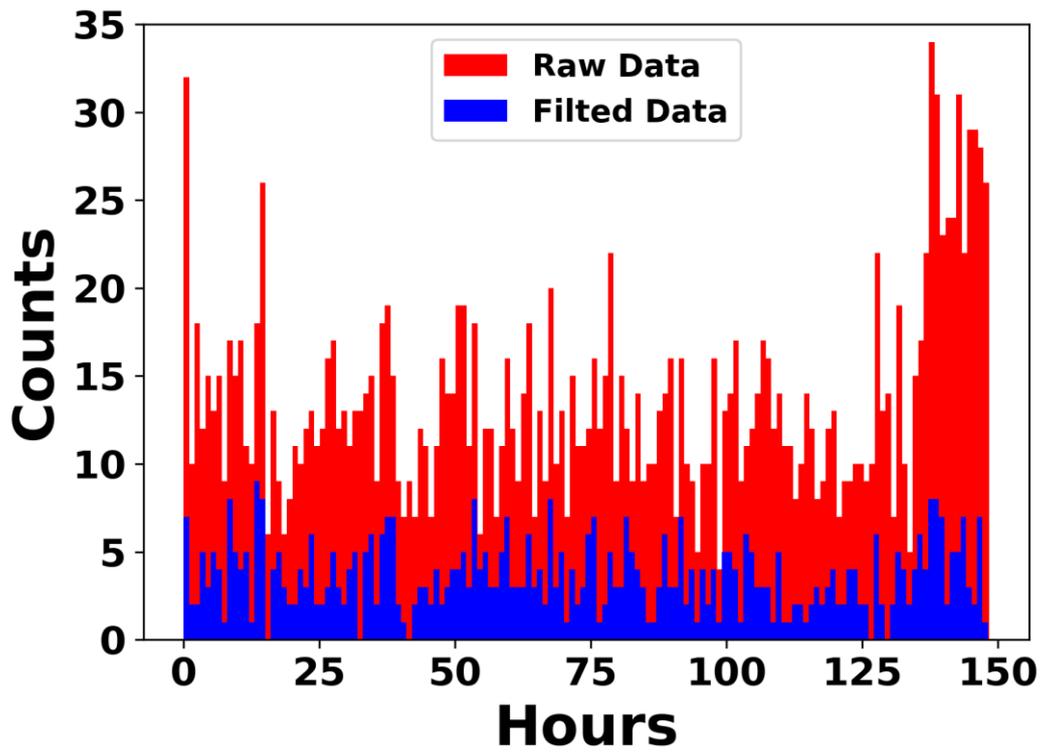

Figure 3. The comparison of background between raw data and filtered data.

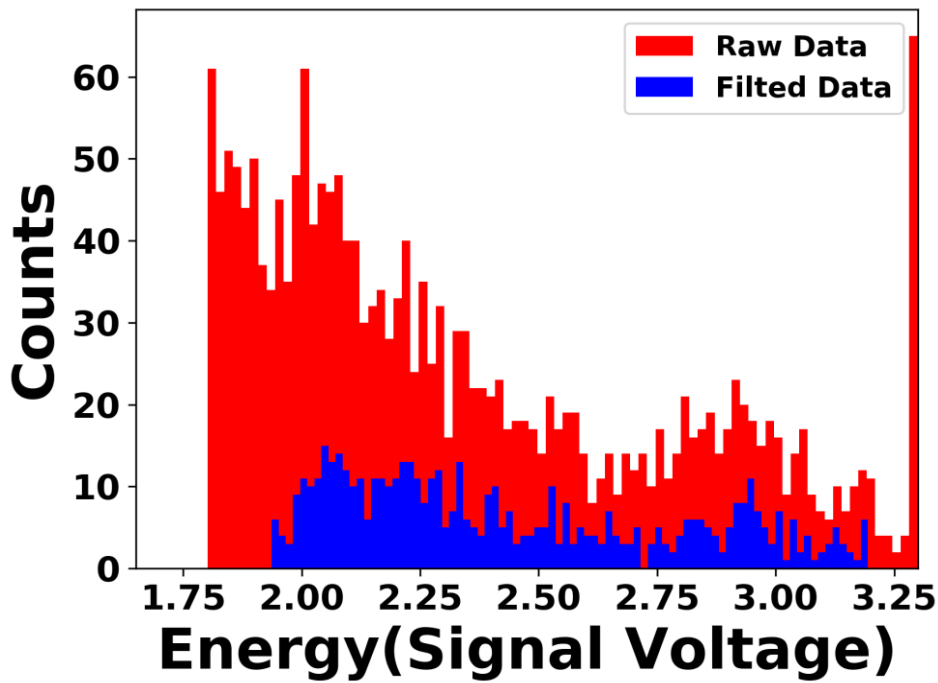

Figure 4. The energy information of background between raw data and filtered data.

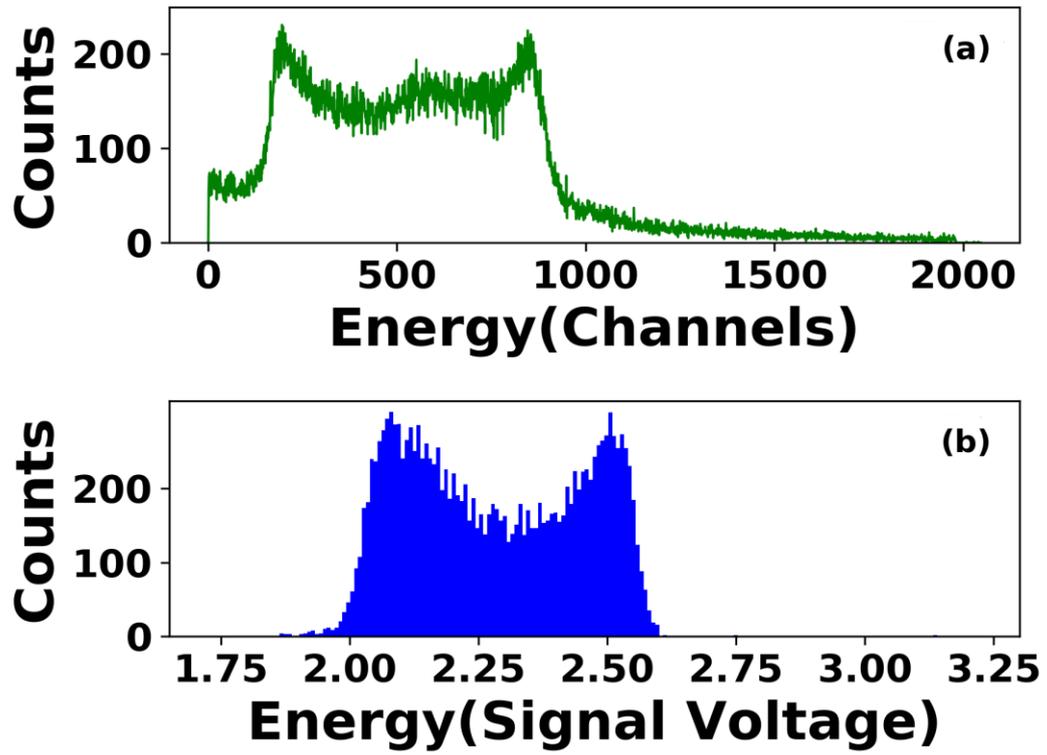

Figure 5. The comparison of energy information between MCA(a) and digital filter method(b).